\documentclass[a4paper,reqno,superscriptaddress,nofootinbib, pra, aps,11pt]{revtex4-2}
\usepackage[centertags]{amsmath}
\usepackage{amsfonts}
\usepackage{amssymb}
\usepackage{amsthm}
\usepackage{titlesec}
\usepackage{newlfont}
\usepackage{stmaryrd}
\usepackage{mathrsfs}
\usepackage{mathtools}
\usepackage{euscript}
\usepackage{graphicx}
\usepackage{enumerate}
\usepackage{todonotes}
\usepackage{color}
\usepackage{float}
\usepackage{orcidlink}
\usepackage{subfigure}
\usepackage{silence}
\usepackage{tikz}
\usepackage{pgf}
\usetikzlibrary{positioning,fit,calc}
\usetikzlibrary{arrows,automata}
\usepackage{wrapfig}
\usepackage{amscd}
\usepackage{cancel}
\usepackage{hhline}
\usepackage{import}

\usepackage{changes}
\date{\today}

\theoremstyle{plain}

\theoremstyle{definition}

\theoremstyle{remark}

\newcommand{\commentA}[1]{[$\mathbb{A}$:{\ {\bf #1}}]}

\DeclareMathOperator{\arctanh}{arctanh}

\usepackage{xcolor}

\makeatletter
\renewcommand\@makefnmark{\hbox{\@textsuperscript{\normalfont \@thefnmark}}}
\makeatother

\let\oldsqrt\sqrt
\def\sqrt{\mathpalette\DHLhksqrt}
\def\DHLhksqrt#1#2{
	\setbox0=\hbox{$#1\oldsqrt{#2\,}$}\dimen0=\ht0
	\advance\dimen0-0.2\ht0
	\setbox2=\hbox{\vrule height\ht0 depth -\dimen0}
	{\box0\lower0.4pt\box2}}

\newcommand{\id}{\textrm{d}}
\def\R{\mathbb R}
\let\ve=\varepsilon   
\newcommand{\bbR}{{\mathbb R}}

\usepackage{url}
\usepackage{lipsum}
\usepackage{mathtools}
\usepackage{lmodern}
\usepackage{anyfontsize}

\begin{document}

\title{Entropy as Noether charge for quasistatic gradient flow}
\author{Aaron Beyen \orcidlink{0000-0002-4341-7661}}

\author{Christian Maes \orcidlink{0000-0002-0188-697X}}

\affiliation{Department of Physics and Astronomy, KU Leuven}

\begin{abstract}
Entropy increase is fundamentally related to the breaking of time-reversal symmetry. By adding the `extra dimension' associated with thermodynamic forces, we extend that discrete symmetry to a continuous symmetry for the dynamical fluctuations around  (nonlinear) gradient flow.  The latter connects macroscopic equilibrium conditions upon introducing a quasistatic protocol of control parameters.  The entropy state function becomes the Noether charge.  As a result, and following ideas expressed by Shin-ichi Sasa and co-workers, the adiabatic invariance of the entropy, part of the Clausius heat theorem, gets connected with the Noether theorem. 
\end{abstract}

\maketitle

\section{Introduction}
Traditionally, within the context of macroscopic physics, gradient flow describes the relaxation to equilibrium.  It appears in a wide variety of contexts, ranging from kinetic theory and hydrodynamics, over Landau theory to machine learning.  Generally speaking, gradient flow is a dynamics of steepest descent where a cost function ({\it e.g.} in the form of a thermodynamic potential) is decreased in updating a physical condition represented as a point on a differential manifold. {\it Quasistatic} gradient flow connects a sequence of equilibrium conditions by slow variation of external parameters ({\it e.g.} pressure or temperature) in the cost function.  Associated with such a reversible evolution appears a state function, the entropy or, more generally, a free energy. 
In particular, the change of entropy is null for a quasistatic adiabatic transformation, \cite{callen1991thermodynamics}.  That statement is part of the Clausius heat theorem, \cite{Campisi_2010}.\\
Reflections on heat and entropy indeed started with the work of Ludwig Boltzmann and Rudolf Clausius, motivated by the First Law and, to a large extent, by the bicentennial work of Sadi Carnot, \cite{Boltzmann_1866, Boltzmann_1871, clausius1871, clausius21871, carnot1824reflexions, thermo_kepler}.  Another important ingredient there is the (generalized) Helmholtz theorem, \cite{Campisi_2010}. For more recent work shedding light on derivations of the (first part of the) Clausius heat theorem, we include the references \cite{thermo_kepler, jonalasinio2023clausius, Saito_2011, prigo, heatconduction, Komatsu_2010, Bertini_2012, clausiusstationarystates, Maes_2015}. 
Connecting that heat theorem to the Noether theorem first appeared in the context of black hole thermodynamics by Wald \textit{et al.}, see \cite{wald_original, wald2, compère2019advanced}.\\

In the present paper, we follow the footsteps of the publications \cite{Sasa_original, Sasa_quantum, langevin_noether, contactgeometry} to understand this `conservation of entropy' from Noether's theorem, \cite{Noether1918}. 
Note that we do not start here from a microscopic dynamics. After all, entropy is fundamentally a \textit{macroscopic} quantity such that it cannot be -directly- connected to Noether's formalism for a mechanical \textit{microscopic} action. Instead, a physically well-motivated coarse-graining approach is required, which defines the level of description (of gradient flow) we are working with in the present paper.  However, related to the deep connection between entropy and fluctuations,  the more powerful view is still to regard the gradient flow as the typical (overwhelmingly plausible in the macroscopic limit) evolution among a much larger collection of trajectories.  That can be formalized via the idea of a path-space action $\mathcal{A}$, a functional defined on all possible trajectories $\Gamma$ with $e^{-N\,\cal A[\Gamma]}$ giving the probability of $\Gamma$, with $N \uparrow \infty$ the huge number of particles or volume elements in the system. In other words,   our point of departure is the structure of dynamical large deviations   (of Markov processes)   around macroscopic evolutions of gradient type.  There, following detailed balance, change of entropy appears as the source term of time-reversal breaking in the ``Lagrangian'' (which is how we call the integrand in the time-integral specifying  $\mathcal{A}$).  Our focus is then on establishing a {\it continuous symmetry} for the quasistatic action and to have the entropy to appear as Noether charge.\\
New in our approach is to give a unifying structure for the above scenario within the context of (nonlinear) gradient flows, \cite{gradientflowmetricspace, ottinger2005beyond,pons, MPR13,patterson2023variational}, which automatically includes aspects of {\it{thermodynamic consistency}}, \cite{Sasa_original}. In addition, it extends the analysis to the macroscopic dynamics of Markov jump models, which did not appear earlier for that relation between entropy and Noether charge.   Furthermore, we discuss the continuous symmetry both in the Lagrangian and in the Hamiltonian framework.    We add several examples to illustrate the main concepts and ideas. \\

The \underline{plan of the paper} is as follows. In Section \ref{section gradient flow}, we review elements of (nonlinear) gradient flow dynamics and we discuss some simple examples. Section \ref{exal} contains the simplest examples.\\
The full strategy and result of the paper are introduced with the Onsager example in Section \ref{liin}.\\ 
The heart of the matter is the large deviation structure, which is recalled in Section \ref{subsection lagrangian and hamiltonian}.  It exposes the canonical structure of the large deviation action for gradient flow, \cite{pons, MPR13}. 
\\
Section \ref{section quasistatic analysis} generalizes that discussion to gradient flow subject to quasistatic protocols in which control parameters are slowly changing in time. The corresponding zero-cost flow is a small perturbation of the instantaneous equilibrium solution. 
Finally, Section \ref{section symmetry} deals with the continuous symmetry of the quasistatic large deviation action for which entropy is the Noether charge.\\
Appendix \ref{section curie weiss} illustrates the procedure for the Curie-Weiss relaxational dynamics.

\section{Gradient flow}\label{section gradient flow}
Gradient flow refers to a certain structure in the dynamics of relaxation to equilibrium, \cite{gradientflowmetricspace,ottinger2005beyond,pons, MPR13}. It is a dynamics of steepest descent where a cost function (or, free energy) is extremized. Gradient flow systems are numerous in physics, ranging from the Master equation for Markov jump processes to the spatially-homogeneous Boltzmann equation on the space of probability measures, \cite{boltzmanngradient}.   We start with a couple of simple examples to illustrate the structure.  

\subsection{Examples}\label{exal}
  
\subsubsection{Linear flow: type-B dynamics}\label{tb}

A standard example of (linear) gradient flow for a scalar density $\rho$   in a regular volume   $\Lambda\subset \R^3,$ with fixed conditions $\rho = \bar{\rho}$ on the boundary $\partial \Lambda$, is given by
\begin{equation}\label{he}
\dot{\rho}=\nabla\cdot\left(\chi(\rho)\nabla \left(\frac{\delta{\cal F}[\rho]}{\delta \rho}\right)\right)
\end{equation}
in which the gradient structure is already explicit.   It can be written as  
  a continuity equation of the form
\begin{align}\label{cequ}
&\dot\rho = - \nabla \cdot j_\rho, \ \text{ where }  \\
&   j_\rho
= \chi\left(\rho\right)\, F,\quad F
= - \nabla \,\mu,\quad \mu = \frac{\delta{\cal F}[\rho]}{\delta \rho}   \label{lmu}
\end{align}

We say that this example is ``linear'' because the current is proportional to the thermodynamic force $F$, for   a   local chemical potential $\mu$ which is the variational derivative of a given free energy functional ${\cal F}[\rho]$.\\ 

That example includes the (even more linear) Smoluchowski equation when considering independent particles, each with position $x$ subject to the overdamped dynamics
\[
\dot x = -\gamma \,\nabla U(x) + \sqrt{2\gamma\,T}\,\xi
\]
with confining potential $U(x)$ on $\bbR^3$ and for standard white noise $\xi$.
Their density $\rho$ (in the limit of $N$ identical independent copies) satisfies the Smoluchowski equation
\begin{equation}\label{smo}
\dot\rho = - \nabla \cdot j_\rho, \quad j_\rho = -\gamma\nabla U\rho -\gamma T\,\nabla\rho 
\end{equation}
for $\chi(\rho) = \gamma \,\rho$ and ${\cal F}(\rho) = \int \id V \ \rho \ U + T\int \id V \ \rho \log\rho$ on some fixed connected volume.  \\

Equations \eqref{he} and \eqref{smo} are of the form
\begin{equation}\label{grade}
\dot z = M\,\id S,\qquad\;\; M = D\, X\, D^\dagger
\end{equation}
for a symmetric positive semi-definite operator $X$, with  $D=-\nabla\cdot$ representing minus the divergence, $D^\dagger=\nabla$ the gradient and $M=-\nabla\cdot\chi\nabla$ for a density-dependent mobility $\chi$ (being a 3 × 3 symmetric non-negative matrix).  The state function $S(z)$ is an entropy for a closed and isolated system, but it becomes some other thermodynamic potential depending on the context for open systems.  In \eqref{he}, the state function $S=-{\cal F}$ is minus the free energy functional $\cal F$. \\
In other examples, the operator $D$ can be the identity operator (for a non-conserved variable), or a stoichiometry matrix (for chemical reactive systems) and its adjoint $D^\dagger$ satisfies the relation $a \cdot D b = b \cdot D^{\dagger} a$. \\
In \eqref{grade}, $\id S$ denotes the derivative with respect to $z$. It is increasing in time along a trajectory $z_t$:
\begin{equation*}
    \dot{S}(z_t) = \id S(z_t) \cdot \dot{z}_t = \id S(z_t) \cdot M \id S(z_t) \geq 0
\end{equation*}
where we used that $M$ is positive semi-definite. The equilibrium state $z^*$ is then defined as the one that maximizes $S$ and pays a central role in the rest of the paper.
Mathematically it makes sense to think of $z$ as a point of a Riemannian manifold, moving
by steepest descent of the functional $-S$. The descent is measured in a metric provided by 
the operator $X$, which is in general related to the physical mobility.\\

This paper uses an extension of gradient flow \eqref{grade} to nonlinear evolutions \cite{genericmaes, gradientflowmetricspace, Kraaij_2020, pons}, which appear {\it e.g.} in the context of Markov jump processes where the relation between force and current is indeed nonlinear. The next example of the Curie-Weiss (mean-field Ising) model is of that type.

\subsubsection{Nonlinear flow: Curie-Weiss dynamics}\label{cwex}

A Curie-Weiss relaxational dynamics   for   the magnetization $m \in [-1,1]$ is given by the evolution
\begin{align}\label{cw m dynamics}
    \dot{m} = 2 \nu  \sqrt{1-m^2} \,\sinh\left(-\beta {\cal F}'(m) \right), \qquad \beta {\cal F}(m) = - \beta \phi(m) - s(m)
\end{align}
for some rate $\nu>0$, inverse temperature $\beta$, energy $ - \phi(m)$, and mixing entropy 
\begin{equation}
    s(m) = - \left[ \frac{1+m}{2} \log\left(\frac{1+m}{2} \right) + \frac{1-m}{2} \log \left(\frac{1-m}{2} \right) \right]
\end{equation}
 The prefactor $\chi(m) = 2 \nu \sqrt{1-m^2} \geq 0$ decides the rate at which the Landau free energy function ${\cal F}$ is decreasing in time   along a curve $m_t$ solving \eqref{cw m dynamics}  :
 \begin{equation*}
     \frac{\id {\cal F}}{\id t}(m_t) = {\cal F}'(m_t) \dot{m}_t = \nu\,{\cal F}'(m_t) \chi(m_t) \sinh\left(-\beta {\cal F}'(m_t) \right) \leq 0
 \end{equation*}
That example stands for a larger class of mean-field systems; see {\it e.g.}, \cite{KMK,Golse_2016}.  We come back to it in Appendix \ref{section curie weiss}. The point here is that \eqref{cw m dynamics} is an example of a nonlinear gradient flow, announcing the general structure introduced in the next section.

\subsection{General Setup}
 The notion of thermal equilibrium requires a physical coarse-graining in terms of a (set of physically well-chosen) macroscopic variable(s).   Similar to \eqref{grade} for the linear gradient flow,   we denote them by $z$ (matter densities $\rho$ in \ref{tb}, magnetization $m$ in Example \ref{cwex},  {\it etc}.) and represent them as points on a smooth manifold which we do not need to make explicit.  We assume that $z$ has even parity under time-reversal,   such that the corresponding current (time-change of $z$ or flux of particle or energy density) is odd.  \\

To describe gradient flow, we need a state function $S_\lambda(z)$ depending on a set of control parameters $\lambda$. Here, $S_\lambda$ might be an entropy, minus a free energy or some other thermodynamic potential depending on the context, {\it etc}.  We write  $ \id S_\lambda$ for the (possibly functional) derivative of $S_\lambda$ with respect to $z$.
The corresponding (nonlinear) gradient flow for $z$ is
 \begin{equation}\label{gger}
\dot z =D j_{z, \lambda}, \qquad j_{z, \lambda} ~ = \partial_f \psi^\star_{  \lambda  }(D^\dagger \id S_\lambda/2;z) 
 \end{equation}
  
 with $D, D^{\dagger}$ as before, and for a convex functional $f \mapsto \psi^\star_{  \lambda  }(f;z)$ where $f$ stands for a thermodynamic force (cotangent vector).  Its (Fréchet) derivative with respect to $f$ is denoted as $\partial_f \psi^\star_{  \lambda  }$ and gives the current $j_{z, \lambda}$.   We refer to  \cite{pons, O,O2, MPR13, genericmaes, ottinger2005beyond} for more details.   
\\ In Example \ref{cwex}, the operators $D = D^{\dagger}$ are the identity, the states $z$ are the values of the magnetization $m\in [-1, +1]$ and   the evolution \eqref{cw m dynamics} can be written as 
\begin{eqnarray*}
\dot m &=& \partial_f \psi^\star_{  \lambda  }(S_\lambda'(m)/2;m)\quad \text{ for } S_\lambda(m) = - \beta {\cal F}(m) \;\text{ and } \nonumber\\
\psi^\star_{  \lambda  }(f,m) &=& \nu \sqrt{1-m^2} \left(\cosh(2 f)-1 \right) \nonumber
\end{eqnarray*}
 
We note that the linear version \eqref{grade} (and thus also Example \ref{tb}) can be rewritten  
with $\psi^\star_{  \lambda  }(f;z) = f\cdot X_\lambda f$ such that $ \partial_f \psi^\star_{  \lambda  }(f;z) = 2X_\lambda f $ for a symmetric positive semi-definite operator $X_\lambda$. 
 

In general, the functional $\psi^\star_{  \lambda  }$ should satisfy the following conditions to yield a meaningful gradient flow, \cite{pons, MPR13}:
\begin{itemize}
    \item min$_f$ $\psi^\star_{  \lambda  }(f;z) = \psi^\star_{  \lambda  }(0;z) = 0$ 
    \item $\psi^\star_{  \lambda  }(f;z)$ is differentiable and \textit{strictly} convex in its first argument, {\it i.e.}, 
    \begin{equation*}
        \psi^\star_{  \lambda  }(t f + (1-t) g;z) < t \psi^\star_{  \lambda  }(f) + (1-t) \psi^\star_{  \lambda  }(g) \quad \forall f,g, \quad t \in (0,1)
    \end{equation*}
    \item $\psi^\star_{  \lambda  }$ is symmetric in $f$, {\it i.e.} $\psi^\star_{  \lambda  }(-f;z) = \psi^\star_{  \lambda  }(f;z)$
\end{itemize}
It then follows that
\begin{align}
 \psi^\star_{  \lambda  }(f;z)\geq & \ \psi^\star_{  \lambda  }(0;z)=0 \quad \forall f, \qquad\text{ and }
\label{psi derivative around 0} \\
 \partial_f \psi^\star_{  \lambda  }(f;z) &= 0 \iff f = 0 \label{cond:psi*-min-at-zero}
\end{align}
{\it i.e.}, $f = 0$ is the \textit{unique} minimizer of $\psi^\star_{  \lambda  }(f;z)$. Moreover, as a consequence of \eqref{psi derivative around 0}, under   a curve $z_t$ satisfying   the dynamics \eqref{gger},
\begin{equation}\label{ly}
  \frac{\id}{\id t} S_{ \lambda  }  (z_t) =\id S_{  \lambda  }(z_t) \cdot \dot z_t=2~D^\dagger\id S_{  \lambda  }(z_t)/2 \cdot \partial_f \psi^\star_{  \lambda  }( D^\dagger \id S_{  \lambda  }(z_t)/2;z_t) \geq 0,  
\end{equation}
which implies the monotonicity of $S_{  \lambda  }(z)$ in time.   
The (unique) maximum $z^*_\lambda$ of $S_\lambda$ (for a given $\lambda$) has $j_{z^*_\lambda} =0$ and represents the unique \underline{equilibrium state},
\begin{equation}\label{eq condition}
\frac{\partial S_\lambda}{\partial z} (z^*_\lambda) = 0, \qquad  S_\lambda(z) \leq   S_\lambda(z^*_\lambda), \qquad \frac{\partial S_{\lambda}}{\partial \lambda}(z_\lambda^*) = 0 
\end{equation}
The last condition indicates that the thermodynamic force conjugate to $\lambda$ needs to vanish in equilibrium; a condition on the $\lambda-$dependence. \\

The original references for the above-sketched framework are \cite{ottinger2005beyond, pons,  MPR13}.  A less trodden path is to consider gradient flow for quasistatic transformations between equilibrium conditions, to which we come in Section \ref{section quasistatic analysis}. 
  That requires time-dependent control parameters $\lambda \to \lambda(t)$, and gets illustrated first in the next section.\\

\section{Onsager example}\label{liin}
The present section wishes to introduce the main concepts and steps 
within the simplest possible Onsager and Onsager-Machlup setups, \cite{O,O2,onsagermachlup}.  It starts from a stochastic dynamics that describes the fluctuations around a macroscopic evolution.\\  
We have one macroscopic variable $z\in \bbR$ that is subject to a fluctuating relaxational dynamics for times $t\in [t_1,t_2]$,
\begin{equation}\label{mz}
\dot z = \gamma \frac{\id S_\lambda}{\id z}  (z)  
+ \sqrt{\frac{2\gamma}{N}}\xi
\end{equation}
  with    ``entropy''  $S_\lambda(z)$ ,    Onsager coefficient $ \gamma > 0$ ,   standard white noise $\xi$ and   a large parameter $N$ such as the number of components or the size of the system in terms of a typical intercomponent distance.   The fact that we call $S_\lambda$ an entropy may refer to the case where $z$ denotes a time-dependent volume with thermally isolated walls subject to a certain imposed pressure $\lambda$.  That is then basically the example of \cite{langevin_noether}. Other scenarios are possible where $S_\lambda$ would for instance stand for minus a free energy instead.\\
 
Equation \eqref{mz} is not (yet) a gradient flow due to the noise term, but becomes one in the macroscopic limit $N \uparrow \infty$ 
 \begin{equation}\label{macroscopic eq}
\dot{z} = \gamma \frac{\id S_\lambda}{\id z} =  D \ \partial_f \psi^\star_{  \lambda  }\left(D^{\dagger}\frac{\id S_\lambda}{\id z}\Big/2 ;z \right), \qquad \text{ with } \  D = 1, \qquad \psi^\star_{  \lambda  }(f;z) = \gamma f^2
\end{equation}

\subsection{Action, Lagrangian and Hamiltonian}\label{alh}

Using the fact that $\xi$ in \eqref{mz} has Gaussian statistics with unit variance, one easily obtains that the   
probability $\mathbb{P}[\Gamma]\propto e^{-N{\cal A}(\Gamma)}$ of a trajectory $\Gamma = ((z_t,\dot z_t), t\in [t_1,t_2])$ is governed by the \textbf{path-space action}
\begin{equation}\label{act}
\cal A[\Gamma] = - S_{  \lambda  }(z_{t_1}) + \int_{t_1}^{t_2}\id t \,L_{  \lambda  }(\dot z_t; z_t)
\end{equation}
with ``Lagrangian,''
\begin{equation}\label{lala}
L_{  \lambda  }(\dot z; z) = \frac 1{4\gamma}\left(\dot{z} - \gamma\frac{\id S_\lambda}{\id z}(z)
\right)^2  \geq 0
\end{equation}
  In \eqref{act}, we assume that   the initial value $z_{t_1}$ gets sampled from the equilibrium distribution $\propto e^{NS_{  \lambda  }(z_{t_1})}$,   while the Lagrangian characterizes the subsequent dynamics. \\
In this framework, the macroscopic equation of motion \eqref{macroscopic eq} corresponds to the trajectory that dominates the path probability $\mathbb{P} \propto e^{- N \mathcal{A}}$ in the limit $N \uparrow \infty$, {\it i.e.}, it corresponds to solving $L(\dot z,z)=0$ for $\dot{z}$ (\textbf{zero-cost flow}). Since the Lagrangian \eqref{lala} is positive and strictly convex in $\dot{z}$, that also amounts to minimizing the Lagrangian \eqref{lala} at fixed $z$, 
\begin{equation}\label{partial l dot z = 0}
    0 = \frac{\partial L_{  \lambda  }}{\partial \dot{z}} = \frac{1}{2 \gamma} \left(\dot{z} - \gamma \frac{\id S_\lambda}{\id z} \right)
\end{equation}

We observe that if \eqref{macroscopic eq} is satisfied, then also the Euler-Lagrange equation for \eqref{lala} is verified
\begin{eqnarray}\label{linear example euler lagrange equation}
0 = \partial_z L_{  \lambda  } - \frac{\id}{\id t} \left(\partial_{\dot{z}} L_{  \lambda  } \right) \iff \ddot{z}  = \gamma^2\,\frac{\id S_\lambda}{\id z}(z)\,\frac{\id^2 S_\lambda}{\id z^2}(z)
\end{eqnarray}
but not \textit{vice versa}. For what follows, Eq.~\eqref{linear example euler lagrange equation}  is not essential, but its importance in other settings such as for using Hamilton-Jacobi theory is clear, \cite{M, F, T}, partially motivating the terminology of Lagrangian for the integrand \eqref{lala}. \\
For completeness, we mention here also that the minimization or variational principle for determining the typical and correct evolution can also be obtained from entropy and dissipation principles, at least for quadratic Lagrangians (Gaussian-type processes) such as \eqref{mz}. That is done via the Onsager dissipation functions or the Rayleighian and it amounts to a form of minimum or maximum entropy production principle; see e.g. \cite{Maesring, mepp, Bruers_2007}. 

As explained in \cite{largedeviationtheory, softmatterbook}, the Rayleighian corresponding to the dynamics \eqref{mz} equals
\begin{equation*}
   R_{  \lambda  }(z, \dot{z}) = \frac{1}{2 \gamma}  \dot{z}^2 - \frac{\id  S_\lambda}{\id  z} \dot{z} = \Phi(\dot{z}) - \dot{S}_\lambda, \qquad \Phi(\dot{z}) = \frac{1}{2 \gamma} \dot{z}^2  
\end{equation*}
Following Onsager's principle, this function is minimized with respect to $\dot{z}$ at fixed $z$, {\it i.e.}, $\frac{\delta R_{  \lambda  }}{\delta \dot{z}} = 0$, to obtain the equations of motion. Here, $\Phi(\dot{z})$ is called the dissipation function since it equals half of the work done to the surroundings per unit time. The Lagrangian $L$ has a similar form to the Rayleighian $R$
\begin{align*}
L_{  \lambda  }(\dot z; z) &= \frac 1{4\gamma}\left(\dot{z} - \gamma\frac{\id S_\lambda}{\id z}(z)
\right)^2  = \frac{1}{4 \gamma} \dot{z}^2 - \frac{1}{2} \frac{\id S_{\lambda}}{\id z}(z) \dot{z} + \frac{\gamma}{4} \left(\frac{\id S_{\lambda}}{\id z}(z) \right)^2 \\
& = \Phi_L(\dot{z}) - \dot{S}_{\lambda} + \frac{\gamma}{4} \left(\frac{\id S_{\lambda}}{\id z}(z) \right)^2
\end{align*}
The last term $ \frac{\gamma}{4} \left(\frac{\id S_{\lambda}}{\id z} \right)^2$ is not present in the Rayleighian, but since we minimise both at fixed $z$, they yield the same gradient flow equation. Consequently, the Lagrangian in the large deviation setup is similar, but not equivalent, to the  Rayleighian in Onsager's principle and both lead to the same equations of motion.
\\
However, we want to emphasize that dynamical large deviation theory is broader and can also be applied to {\it e.g.} discrete Markov jump processes \cite{lagrangianmarkov} and nonquadratic Lagrangians (see Section \ref{subsection lagrangian and hamiltonian}), which do not fall under Onsager's principle. Moreover, the path-space action is not only used to derive the equations of motion but also allows calculating stationary probability densities \cite{M, F, T, maesresponse}, {\it etc}. \\

The Legendre transform of $L_{  \lambda  }(z,\dot z)$  is the ``Hamiltonian,''
\[
H_{  \lambda  }(f;z) =  \sup_{\dot{z}} \left\{ f \dot{z} - L_{  \lambda  }(\dot{z}; z) \right\}  = \gamma\,f^2 + \gamma f\,\frac{\id S_\lambda}{\id z} = \gamma\,f\,\left(f+ \frac{\id S_\lambda}{\id z}\right)
\]

The variable $f$ is dual to the current $\dot{z}$, and represents a \textbf{thermodynamic force}. As \eqref{macroscopic eq} corresponds to the condition $L(\dot{z}; z) = 0$, it follows from the Legendre transform (see Section \ref{subsection lagrangian and hamiltonian}) that the \textbf{Hamiltonian zero-cost flow} becomes
\begin{equation}\label{zero cost flow onsager ham}
\dot{z} = \partial_{f}H_{  \lambda  }(f = 0;z) 
= \gamma\, \frac{\id S_\lambda}{\id z}(z)
\end{equation}
which indeed gives the correct result. If this equation is satisfied, then also the more general Hamilton's equations of motion are verified
\begin{eqnarray}
    \frac{\partial H_{  \lambda  }}{\partial f}(f;z) = \dot z &=& 2\gamma f + \gamma\frac{\id S_\lambda}{\id z}(z)
    \label{feq}\\
    -\frac{\partial H_{  \lambda  }}{\partial z}(f;z) = \dot{f} &=& \gamma f\, \frac{\id^2 S_\lambda}{\id z^2}(z) \label{haha}
\end{eqnarray}
for $f \equiv 0$, but not {\it vice versa}.

\subsection{Quasistatic analysis}
 
We introduce a \textbf{slow time-dependence} in the control parameter $\lambda \to \lambda(\varepsilon t) =  \lambda_0   \,\cos(\varepsilon t)$ for a small rate $\ve >0$.  For simplicity of illustration, we put $t_1=0, t_2=\tau/\ve$   in \eqref{act}  , and choose as `entropy' function
\begin{equation}\label{entropy function example}
     S_{\lambda(\ve t)}(z)   = - \frac 1{2}\big(z-  \lambda_0   \cos(\ve t) \big)^2 
\end{equation}
where the instantaneous equilibrium state follows
\[
 \,z_{\lambda(\ve t)}^* = \lambda(\ve t) = \lambda_0   \cos(\ve t)
\]
  The choice of entropy \eqref{entropy function example} looks arbitrary (as it is an example) but in fact, it represents the simplest setup, close-to-equilibrium, that allows deforming the equilibrium state slowly and doing an exact computation. \\

We solve the macroscopic equation
\begin{equation}\label{qsm}
\dot z = \gamma\frac{\id    S_{\lambda(\ve t)}  }{\id z}(z) = -\gamma \big(z-   \lambda_0   \cos(\ve t) \big),\quad t\in [0,\tau/\ve] 
\end{equation}
which we start from equilibrium  $z_0 =   \lambda_0$ with maximum entropy $S_{\lambda_0}(z_0) = 0$,   after which the solution of \eqref{qsm} slowly deviates from it.   An exact calculation yields 
\begin{align*}
    z^{(\ve)}_t & = \lambda_0 \ e^{- \gamma t} + \lambda_0 \frac{\gamma^2}{\gamma^2 + \ve^2} \left( \cos(\ve t) - e^{- \gamma t} \right) +\ve \ \lambda_0\frac{\gamma}{\gamma^2+\ve^2} \sin (\ve t)\\
    \dot z^{(\ve)}_t &= \lambda_0 \frac{\gamma \ \ve^2}{\gamma^2 + \ve^2} \left( \cos(\ve t) - e^{- \gamma t} \right) - \ve \ \lambda_0 \frac{\gamma^2}{\gamma ^2+\ve^2} \sin (\ve t)
\end{align*}
and it solves
\[
   L_{\lambda(\ve t)} (\dot z; z) =  \frac 1{4\gamma}\big[\dot z + \gamma \big(z-   \lambda_0    \cos(\ve t) \big) \big]^2 = 0
\]
which expresses the quasistatic motion \eqref{qsm} as the zero-cost flow for the Lagrangian $   L_{\lambda(\ve t)}  (\dot z;z)$ of the path-space action as introduced in \eqref{act}--\eqref{lala}. Up to order $O(\ve^2$), the solution can be written in the general form
\begin{align*}
    z^{(\ve)}_t & =  \,z_{\lambda(\ve t)}^* + \ve \Delta z_{\ve t} + O(\ve^2)   =   \lambda_0   \cos(\ve t) + \varepsilon \frac{  \ \lambda_0   }{\gamma} \,\sin(\ve t)  + O(\ve^2) \\
    \dot z^{(\ve)}_t &=   \ve \partial_{\lambda} z^*_{\lambda(\ve t)} \dot{\lambda}(\ve t) + O(\ve^2)  = -   \ve \ \lambda_0    \sin(\ve t) + O(\ve^2)
\end{align*}
   
Similarly, the ``quasistatic Hamiltonian'' equals  $   H_{\lambda(\ve t)}  (f,z) = \gamma\, f^2 - \gamma\,f\,\big(z-  \lambda_0    \cos(\ve t) \big)$, and \eqref{qsm} combined with $f=0$ express the zero-cost flow in the corresponding Hamilton equation \eqref{feq}.\\

From substituting the zero-cost flow in \eqref{entropy function example}, it immediately follows that
\begin{equation*}
      S_{\lambda(\ve t)}(z^{(\ve)}_t) = S_{ \lambda_0}  (z_0)   - \varepsilon^2 \frac{  \lambda_0   ^2}{2 \gamma} \sin^2(\varepsilon t) + O(\ve^3)
\end{equation*}
Therefore, trivially here, the 
entropy $   S_{\lambda(\ve t)} $ is conserved 
uniformly in time $t\in [0,\tau/\ve]$, for the zero-cost flow in the quasistatic limit $\ve \downarrow 0$ where we start from equilibrium.  There is no surprise, but based on this observation, the much deeper question arises whether this `conservation of entropy' can be derived from Noether's theorem as well, and the answer is yes.  We continue for the same example  by indicating the argument
for this central claim, that for the action $\mathcal{A}$ in \eqref{act}, one can perform mathematical steps akin to Noether's theorem for more conventional actions, leading to the statement that entropy is the Noether charge for this action and thus conserved \textit{on shell}\footnote{  Solving $L_{  \lambda  }(\dot{z}; z)$ = 0 gives the most likely current when in state $z$ at parameters $\lambda$, and should produce the macroscopic equation \eqref{macroscopic eq}. Calculations are called “on shell” when $\dot{z}$ and $z$ are related in exactly that way.}.

\subsection{Symmetry transformations in Noether's theorem}

We start in the Hamiltonian formalism where we consider a shift in the thermodynamic force $f$ at time $t$ depending on $z_t$,  
\begin{equation}\label{onslow}
f_t\rightarrow f_t + \eta  \,q_{\lambda(\ve t)}(z_t)  
\end{equation}
We want to see the change in the action to linear order in $\eta$ to understand what function $ \,q_{\lambda(\ve t)}(z)  $ leads to the entropy as Noether charge: 
\begin{align}
    \delta \mathcal{A} &= \int_{0}^{\tau/\ve} \id t \ \left(\eta\,\dot{z}_t  \, q_{\lambda(\ve t)}(z_t)   - \delta  H_{\lambda(\ve t)}(f_t;z_t)   \right) \label{change action ham linear example} \\
    \delta   H_{\lambda(\ve t)}  (f_t;z_t) &=   H_{\lambda(\ve t)}(f_t + \eta q_{\lambda(\ve t)}(z_t) ,z_t) - H_{\lambda(\ve t)}(f_t,z_t) = \eta \  q_{\lambda(\ve t)}(z_t) \partial_fH_{\lambda(\ve t)}(f_t,z_t)    \\
    & = \eta\, \gamma\,  q_{\lambda(\ve t)}(z_t)   \big(2 f_t - z_t +   \lambda_0    \cos(\ve t) \big) \nonumber
\end{align}
Substituting for $(f_t,z_t)$ the zero-cost flow $z^{(\ve)}_t =   \lambda_0    \cos(\ve t) + \varepsilon   \lambda_0   /\gamma \sin(\ve t)  + O(\ve^2)$,  $f_t^{(\ve)} = 0$, we get
\[
\delta  H_{\lambda(\ve t)}  = - \eta ( \ve   \lambda_0   \,\sin(\ve t) + O(\ve^2)) \,q_{\lambda(\ve t)}(z_t^{(\ve)})\qquad t\in [0,\tau/\ve]
\]
The choice 
\begin{equation}\label{f shift onsager}
    \,q_{\lambda(\ve t)} (z) = -(z -   \lambda_0    \cos(\ve t)) = \frac{\id S_{\lambda(\ve t)} }{\id z}(z)
\end{equation}
makes $\delta  H_{\lambda(\ve t)}  = O(\varepsilon^2)$ for all $t\in [0,\tau/\ve]$
which implies that the Hamiltonian and its time-integral over $[0,\tau/\ve]$ in \eqref{change action ham linear example} are invariant in the quasistatic limit $\ve \downarrow 0$ under the zero-cost flow. Hence, 
 
\begin{equation*}
    \delta \mathcal{A} = \eta\,\int_{0}^{\tau/\ve} \id t \left[ \frac{\id  S_{\lambda(\ve t)} }{\id z}(z_t^{(\ve)}) \dot{z}_t^{(\ve)} + O(\ve^2) \right] = \eta\,\int_{0}^{\tau/\ve} \id t \left[ \frac{\id  S_{\lambda(\ve t)} }{\id t}(z_t^{(\ve)}) - \ve \frac{\partial  S_{\lambda(\ve t)} }{\partial \lambda}(z_{\lambda(\ve t)}^*) \dot{\lambda}(\ve t) +  O(\ve^2) \right] 
\end{equation*}
Following the equilibrium conditions \eqref{eq condition}, one has $\partial_\lambda S_\lambda(z^*_\lambda) = 0$ such that
 
the change in the action is given by a total derivative in the quasistatic limit
\begin{equation}\label{quasisymmetry linear example}
    \delta \mathcal{A} = \eta\,\int_{0}^{\tau/\ve} \id t \left[ \frac{\id S_{\lambda(\ve t)}}{\id t} + O(\ve^2) \right] = \eta\,\left(S_{\lambda_f}(z_{t_2}^{(\ve)}) - S_{\lambda_i}(z_{t_1}^{(\ve)}) + O(\ve)\right)
\end{equation}
In other words, the change in the action under the continuous symmetry $f\rightarrow f + \eta \,\frac{\id  S_{\lambda(\ve t)} }{\id z}$ gives the change in entropy   in the quasistatic limit $\ve \downarrow 0$. Moreover, 
the change in the action under the zero-cost flow can also be written as
\begin{align*}
    \delta \mathcal{A}|_{\text{on shell}}  &=  \int_{0}^{\tau/\ve} \id t \left[\eta \dot{z}_t^{(\ve)} \  q_{\lambda(\ve t)}(z_t^{(\ve)}) - \delta H_{\lambda(\ve t)}(f_t^{(\ve)}; z_t^{(\ve)})  \right] \\
    & = \eta \int_{0}^{\tau/\ve} \id t \ q_{\lambda(\ve t)}(z_t^{(\ve)}) \left[\dot{z}_t^{(\ve)} - \partial_f H_{\lambda(\ve t)}(0; z_t^{(\ve)})  \right] = 0
\end{align*}
where we used \eqref{zero cost flow onsager ham} in the last line, implying the stationarity of the action. Combining this with \eqref{quasisymmetry linear example}, it thus follows   that the entropy corresponds to the Noether charge of a shift in $f$ by  $\eta\,\id S_{  \lambda(\ve t)  }/\id z$ when the variables are following the quasistatic zero-cost flow \eqref{qsm}   and in the quasistatic limit $\ve \downarrow 0$.     That is exactly the result found by Sasa \textit{et al.} in \cite{langevin_noether} for the Langevin equation \eqref{mz}. In what follows, we show that the same can be proven in the Lagrangian formalism and also works for the more general nonlinear gradient flow dynamics \eqref{gger}.  
\\
\\
A similar procedure works in the Lagrangian formalism.  We rewrite the Lagrangian \eqref{lala} as
\begin{equation*}
    L_{\lambda(\ve t)} (\dot z; z) = \frac{\dot{z}^2}{4 \gamma} - \frac{\dot{z}}{2} \frac{\id  S_{\lambda(\ve t)} }{\id z}(z) - \frac{\gamma}{2}  S_{\lambda(\ve t)} (z)
\end{equation*}
using the entropy $ S_{\lambda(\ve t)} $ in \eqref{entropy function example}. The relevant symmetry becomes a shift in the current $j = \dot{z}$, {\it i.e.}, $\dot{z}_t \to \dot{z}_t + \eta \,   p(\dot{z}_t)  $, without changing the macroscopic variable $z$, i.e. $\delta z_t = 0$. As such, the transformations of the generalized `velocities' are not determined by those of the coordinates $z$ and of the time $t$. This kind of transformation is typically not considered for Noether's theorem, as it does not lead to a symmetry of the action. However, it has appeared in \cite{nonconservativenoether, Vittal1988} in describing Noether's theorem for classical mechanical systems subject to nonconservative forces. 
Applying this idea to our setup, one finds to linear order in $\eta$
\begin{align*}
  \delta \mathcal{A} &= \int_{0}^{\tau/\ve} \id t \  \delta  L_{\lambda(\ve t)}(\dot z_t; z_t)  \\
 \delta   L_{\lambda(\ve t)}(\dot z_t; z_t)  &= L_{\lambda(\ve t)} (\dot{z}_t + \eta   p(\dot{z}_t)  ; z_t)   -  L_{\lambda(\ve t)} (\dot{z}_t; z_t) 
 = \eta \frac{\partial  L_{\lambda(\ve t)} }{\partial \dot{z}}   p(\dot{z})   
 \\
     & = \frac{\eta}{2} \left(\frac{\dot{z}}{\gamma} - \frac{\id  S_{\lambda(\ve t)}}{\id z}  \right)   p(\dot{z})  
\end{align*}
Upon choosing
\begin{equation}\label{j shift onsager}
p(\dot{z})   = - 2 \dot{z}    
\end{equation}
this reduces to
\begin{equation*}
    \delta  L_{\lambda(\ve t)}(\dot z_t; z_t)  = \eta \left(\frac{\id  S_{\lambda(\ve t)} }{\id z}(z_t) \ \dot{z}_t - \frac{\dot{z}_t^2}{\gamma} \right) = \eta \left(\frac{\id  S_{\lambda(\ve t)} }{\id t}(z_t)   - \ve \frac{\partial S_{\lambda(\ve t)}}{\partial \lambda}(z_t) \dot \lambda(\ve t)   - \frac{\dot{z}^2}{\gamma} \right)
\end{equation*}
Substituting for $(z_t,\dot{z}_t)$ the zero-cost flow $z^{(\ve)}_t =   \lambda_0    \cos(\ve t) + \varepsilon   \lambda_0   /\gamma \sin(\ve t)  + O(\ve^2)$,  $\dot{z}_t^{(\ve)} = - \ve   \lambda_0   \sin(\ve t) + O(\ve^2)$ and using that  $\partial_\lambda S_\lambda(z_\lambda^*) = 0,$   we get
\begin{equation*}
      \delta \mathcal{A} = \eta \int_{0}^{\tau/\ve} \id t \left[  \frac{\id  S_{\lambda(\ve t)}}{\id t} + O(\ve^2) \right] = \eta \left( S_{\lambda_f}(z_{t_2}^{(\ve)}) - S_{\lambda_i}(z_{t_1}^{(\ve)}) + O(\ve) \right)
\end{equation*}
which is the same result as the Hamiltonian case \eqref{quasisymmetry linear example}.   Moreover, the change in the action under the zero-cost flow can also be written as
\begin{align*}
    \delta \mathcal{A}|_{\text{on shell}} 
    &= \eta \int_{0}^{\tau/\ve} \id t  \ \delta   L_{\lambda(\ve t)}(\dot z_t^{(\ve)}; z_t^{(\ve)}) 
 = - 2 \eta \int_{0}^{\tau/\ve} \id t  \ \dot{z}_t^{(\ve)} \cdot \frac{\partial  L_{\lambda(\ve t)} }{\partial \dot{z}}(\dot z_t^{(\ve)}; z_t^{(\ve)})    = 0   
\end{align*}
where we used \eqref{partial l dot z = 0} in the last line, implying the stationarity of the action. Therefore, the entropy corresponds to the Noether charge of a shift in   the current $j \to j - 2 \eta\ j$   when the variables are following the quasistatic zero-cost flow \eqref{qsm}   and in the quasistatic limit $\ve \downarrow 0$.  \\

The   shifts $f \to f + \eta q_{\lambda(\ve t)}(z) $ and $j \to j + \eta p(j)$,   as used respectively for the Hamiltonian and the Lagrangian, are related   \textit{on shell}   because
\begin{equation*}
    f = \frac{\partial  L_{\lambda(\ve t)}}{\partial \dot{z}} = \frac{\dot{z}}{2 \gamma} - \frac{1}{2} \frac{\id  S_{\lambda(\ve t)}}{\id z} \Longrightarrow   q_{\lambda(\ve t)}(z) = \frac{p(j(z,t))}{2 \gamma} = - \frac{\dot{z}(z,t)}{\gamma} = \frac{\id  S_{\lambda(\ve t)}}{\id z}(z)  
\end{equation*}
where we have used the macroscopic equation \eqref{qsm} in the last line.

 \section{Path-space action}\label{subsection lagrangian and hamiltonian}
 The structure of gradient flow (Section \ref{section gradient flow}), which is a coarse-grained or macroscopic evolution, is intimately related to the structure in its dynamical fluctuations, \cite{pons, MPR13}.  There we enter dynamical large deviation theory, giving the probabilities of trajectories\footnote{   In difference with equilibrium statistical mechanics where, for its static fluctuation theory, one may forget about the microscopic dynamics and the central object is the Hamiltonian, here, one constructs dynamical ensembles of spacetime trajectories, \cite{lagrangianstatmech, lagrangianmarkov}. } for many-particle systems,   \cite{largemarkov, jump_action, lagrangianmarkov} . More specifically, we have in mind a macroscopic system with a large parameter $N$ indicating the number of components or the size of the system in terms of a typical intercomponent distance.  We consider all possible evolutions in a time-interval $[t_1,t_2]$ involving the macroscopic variable $z$ and corresponding current variable $j 
 $, which satisfy $\dot z = Dj$ and that are initially sampled from an equilibrium distribution determined   by the entropy $S_{  \lambda}(z_{t_1})$. The dynamical content of the path-probability is specified by what is called the Lagrangian\footnote{ 
 This Lagrangian does not have the standard kinetic minus potential energy form.  Yet,  the names ``action” and ``Lagrangian” are also justified in analogy with the path-integral approach to quantum mechanics and due to the close analogy between the low-noise limit (large $N$ in {\it e.g.} \eqref{mz} ) of SDEs and the semi-classical or WKB approximation \cite{largedeviationtheory, largedeviationtheory2}.  Note that the Lagrangian depends on $(j,z)$ and does not contain $\dot{z}$ directly unless the operator $D = 1$ in \eqref{gger} such that $\dot{z} = j$. We refer to \cite{largemarkov, jump_action} for more details on this point.  }.  More precisely and as first illustrated in Section \ref{alh}, the Lagrangian $L(j;z)$ of the system governs the path-probabilities via
\begin{equation}\label{large deviation prob z}
    \mathbb{P}\Big[ \{z_t, j(t) \}_{t_1 \leq t \leq t_2}   \Big] \propto \exp \left\{- N \,\left[-S_{  \lambda}(z_{t_1}) + \int_{t_1}^{t_2} \id t \ L_{  \lambda  }\big(j(t); z_t \big)\right]\right\}  
\end{equation}
 in leading order for $N \uparrow \infty$.
  In that limit,   all paths   are   exponentially   suppressed compared to the zero-cost flow.  As is the standard case, we require that $L_{  \lambda  }$ is nonnegative and  (strictly) convex such that   putting $L_{  \lambda  }(j; z) = 0$   yields   a (unique) current   $j = j_{z, \lambda}$, producing   the macroscopic equation of motion \eqref{gger}   at minimal cost/maximal probability. Hence,   obtaining the `equations of motion' or   the optimal path   of the system can be interpreted as an action minimization problem, which is similar to the Least Action Principle in Lagrangian mechanics.   Yet, in a way, it simply amounts to  a nonlinear extension of the Onsager's minimum dissipation principle  mentioned in Section III.A.
\\
\subsection{Lagrangian and Hamiltonian}

We summarize the properties we suppose about the Lagrangian in \eqref{large deviation prob z}; see also \cite{pons, MPR13}:
\begin{itemize}
    \item $ L_{  \lambda  }(j;z)  \geq 0 $ 
    \item $ L_{  \lambda  }(j;z)  \text{ is \textit{strictly} convex in its first argument, {\it i.e.},} \nonumber$
    \begin{equation}
        L_{  \lambda  }(c\, j + (1-c)\, j';z) < c \,L_{  \lambda  }(j;z) + (1-c)\, L_{  \lambda  }(j;z), \quad \forall j,j', \quad c \in (0,1) \label{pL}
    \end{equation}
    \item $L \text{ induces an evolution equation for } z \text{ by requiring zero-cost}:\nonumber $
    \begin{equation}
        \qquad \qquad L_{  \lambda  }(j_{z, \lambda}; z) = 0 \iff \dot{z} = D j_{z, \lambda}, \label{l = 0 condition}
    \end{equation}
     and we refer to calculations as ``on shell'' when $j$  $\text{ and } \dot{z} \text{ are related in that way}$.
\end{itemize}

Therefore, for every $z$, the Lagrangian has a unique minimum $L_{  \lambda  }(j;z) = 0$ at $j = j_{z, \lambda}$, inducing the gradient flow $\dot{z} = D j_{z, \lambda}$. That (first-order) evolution is a distinguished subclass of the Lagrangian flow described by the (second-order) Euler-Lagrange equation
\begin{equation}\label{euler lagrange equation}
    \partial_z L_{  \lambda  }(z_t,j_{z, \lambda}(t)) - \frac{\id}{\id t} \Big( \partial_j L_{  \lambda  }(z_t,j_{z, \lambda}(t)) \Big) = 0,
\end{equation}
 if the initial conditions match.  
To see this, note that when for all $z, L_{  \lambda  }(z,j_{z, \lambda}) = 0$, ($L_{  \lambda  }$ is identically zero on shell), then
\begin{equation}\label{dLdz = 0 on shell}
    0 = \frac{\id}{\id z} \Big( L_{  \lambda  }(z,j_{z, \lambda}) \Big) = \partial_z L_{  \lambda  }(z,j_{z, \lambda}) + \partial_j L_{  \lambda  }(z,j_{z, \lambda}) \ \partial_z j_{z, \lambda}
\end{equation}
Since the Lagrangian is minimal at $j = j_{z, \lambda}$, one has $\partial_j L_{  \lambda  }(z,j_{z, \lambda}) = 0$ such that \eqref{dLdz = 0 on shell} gives $\partial_z L_{  \lambda  }(z,j_{z, \lambda}) = 0$ on shell. As both terms are zero, it immediately follows that also the Euler-Lagrange equation \eqref{euler lagrange equation} is satisfied. Hence,  a solution of the gradient flow $\dot{z} = Dj_{z, \lambda}$ solves the Euler-Lagrange equation as well.   Even though we do not need the second-order dynamics \eqref{euler lagrange equation} in what follows, that result is still interesting and an important check since, {\it a priori}, one expects the Noether charge to only be conserved when the full Euler-Lagrange equations are satisfied. \\


The Lagrangian $L_{  \lambda  }$ is related to  the Hamiltonian $H_{  \lambda  }$ through the usual Legendre
transform,
\begin{equation}\label{hmd}
    H_{  \lambda  }(f;z) = \sup_j \Big\{j \cdot f - L_{  \lambda  }(j;z) \Big\}, \qquad   L_{  \lambda  }(j;z) = \sup_f \Big\{j \cdot f - H_{  \lambda  }(f;z) \Big\}
\end{equation}
where $f$, the variable dual to the current $j$, represents a thermodynamic force. 

The properties \eqref{pL}--\eqref{l = 0 condition} of the Lagrangian translate into
\begin{align}
    & 
    \llap{\textbullet \hspace{10pt}} \text{$L_{  \lambda  } = 0$ at its minimum} \Longrightarrow H_{  \lambda  }(0;z) = \sup_j \Big\{ - L_{  \lambda  }(j;z) \Big\} = -\inf_j \Big\{ L_{  \lambda  }(j;z) \Big\} = 0 \label{h(0,z) = 0} \\
    &\llap{\textbullet \hspace{10pt}}  
    H_{  \lambda  }(f;z) \text{ is \textit{strictly} convex in its first argument} \label{h strictly convex}\\
    &\llap{\textbullet \hspace{10pt}} 
    L_{  \lambda  }(j_{z, \lambda}; z) = 0 \Longrightarrow \sup_f \Big\{j_{z, \lambda} \cdot f - H_{  \lambda  }(f;z) \Big\} = 0 \label{sup L = 0}
\end{align}

The last property can be further simplified by noting from \eqref{h strictly convex} that $j_{z, \lambda} \cdot f - H_{  \lambda  }(f;z) $ is strictly concave in $f$ and thus has at most one maximum. Since $j_{z, \lambda} \cdot 0 - H_{  \lambda  }(0;z) = 0$ due to \eqref{h(0,z) = 0}, the supremum \eqref{sup L = 0} is reached at $f = 0$ only. Remembering that $j_{z, \lambda}$ also satisfies $j = \partial_f H_{  \lambda  }(f;z)$, it follows that the zero-cost flow becomes
\begin{equation}\label{zero cost flow}
     \dot{z} = D j_{z, \lambda} \iff L_{  \lambda  }(j_{z, \lambda}; z) = 0 \iff j_{z, \lambda} = \partial_f H_{  \lambda  }(f;z) \text{ and } f = 0
\end{equation}
The equation $j_{z, \lambda} = \partial_f H_{  \lambda  }(f;z)$ is the analogue of the first Hamilton equation;
the condition $f = 0$ is a subclass of the second Hamilton equation
\begin{equation*}
    \dot{f} = - \partial_z H_{  \lambda  }(f;z)
\end{equation*}
since $\partial_z H_{  \lambda  }(0;z) = 0$ as follows from \eqref{h(0,z) = 0}.
Summarizing, similar to how $L_{  \lambda  }(j_{z, \lambda}; z) = 0$ satisfies the Euler-Lagrange equations, the relations $j_{z, \lambda} = \partial_f H_{  \lambda  }(f;z) \text{ and } f = 0$ solve Hamilton's equations.

\subsection{Detailed balance}
Detailed balance plays an essential role in the Clausius heat theorem, which when combined with quasistatic changes, defines a {\it reversible} evolution.   Assuming that $z$ is even and $j$ odd under kinematic time-reversal\footnote{  We always assume that trajectories have $j$ proportional to $\dot z$.  Physically, for $z$ we have in mind particle, volume or energy densities, while $j$ refers to the corresponding fluxes.  In the case where $z$ stands for a momentum distribution, we have that $z$ is odd and the momentum flux is even under time-reversal.  We do not treat that last case separately but it does not present further challenges.  },
time-reversal invariance is equivalent to 
 
\begin{equation*}
    \int_{t_1}^{t_2} \id t \  \Big[L_{  \lambda  }\left(- j(t); z_{t} \right) - L_{  \lambda  }\left(j(t); z_t \right) \Big] = S_{  \lambda}(z_{t_2})- S_{  \lambda}(z_{t_1}) = \int_{t_1}^{t_2} \id t \frac{\id S_{  \lambda}}{\id t}(z_t) 
\end{equation*}
Since this equation is true for all times $t_1,t_2$, we get detailed balance in the form
\begin{equation}\label{dbe}
    L_{  \lambda  }\left(- j(t); z_t \right) - L_{  \lambda  }\left(j(t); z_t \right) = \frac{\id S_{  \lambda}}{\id t}(z_t) = j(t) \cdot D^{\dagger} \id S_{  \lambda}(z_t)
\end{equation}
indicating that the Lagrangian only changes by a temporal boundary term under time-reversal. Taking the limit $j \to 0$, this relation indicates that the tangent to $L$ at $j = 0$ is given by $-2 D^{\dagger} \id S_{  \lambda}$.

\begin{figure}[H]
    \centering
    \includegraphics[width = 12 cm]{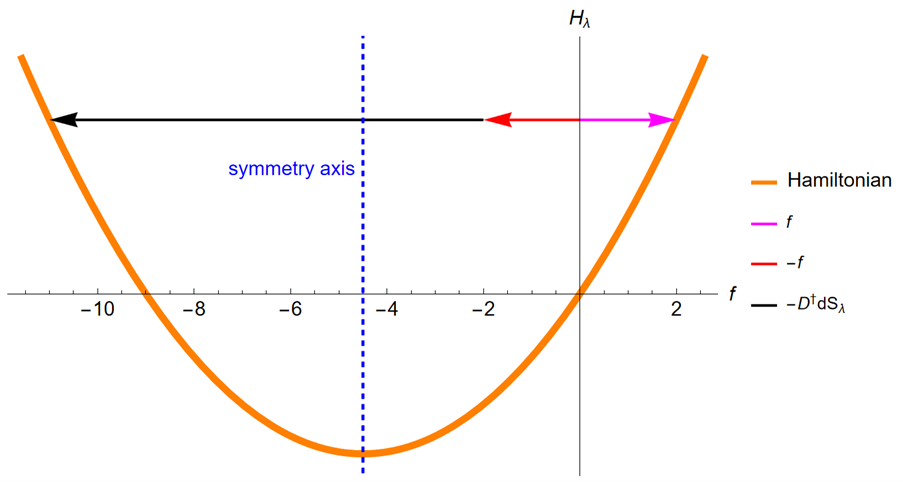}
    \caption{Visualisation of the detailed balance condition \eqref{dbham} for the Onsager example \ref{liin} with $z = -8$. Note the symmetry around the dashed line $f = - D^{\dagger} \id S_{  \lambda}/2$.}
    \label{fig ham symmetry}
\end{figure}

Similarly, in the Hamiltonian formalism, using \eqref{hmd}, the detailed balance condition becomes
\begin{equation}\label{dbham}
    H_{  \lambda  }(f;z) = H_{  \lambda  }(-f - D^{\dagger} \id S_{  \lambda}; z)
\end{equation}
{\it i.e.}, that the Hamiltonian is symmetric in $f$ around the point $- D^{\dagger} \id S_{  \lambda}/2$, as indicated in Fig. \ref{fig ham symmetry}. Moreover, it indicates that the discrete time-reversal symmetry can be understood as a shift $f\rightarrow f - (2f + D^\dagger dS_{  \lambda})$ in the thermodynamic force.  That observation plays center stage in our application of (the Hamiltonian version of) Noether's theorem in Section \ref{section quasistatic analysis} and has already appeared in Example \ref{liin}. 

\subsection{Canonical structure}\label{canonical structure}
As shown in \cite{pons, MPR13, genericmaes}, the path-space action generates the gradient descent dynamics \eqref{gger} under the detailed balance condition \eqref{dbe} combined with the properties of the Lagrangian \eqref{l = 0 condition}.  We briefly recall the argument.\\

It is the frenetic part in the path-space action ({\it i.e.}, the time-symmetric part, \cite{fren}) that determines the structure \eqref{gger}.
To start, we write
\begin{equation}\label{low}
\frac{L_{  \lambda  }(j;z) + L_{  \lambda  }(-j;z)}{2} = \psi_{  \lambda  }(j;z) +\psi^\star_{  \lambda  }(D^\dagger \id S_{  \lambda}/2;z)
\end{equation}
where \begin{equation}\label{pst}
\psi^\star_{  \lambda  }(f;z) = \sup_j\left[j\cdot f - \psi_{  \lambda  }(j;z)\right]
\end{equation} 
is the Legendre transform of
\begin{eqnarray}
\psi_{  \lambda  }(j;z) &=& \frac{1}{2}\bigl[L_{  \lambda  }(-j;z) + L_{  \lambda  }(j;z)\bigr]-L_{  \lambda  }(0;z) \nonumber\\
 &=& \frac{1}{2}\,j\cdot D^\dagger \id S_{  \lambda}(z) + L_{  \lambda  }(j;z) -L_{  \lambda  }(0;z)  \label{Lpsi}
\end{eqnarray}
The second equality follows from detailed balance \eqref{dbe} and implies that $\psi_{  \lambda  }(j;z)$ is convex in $j$.
From the first line, we see that $\psi_{  \lambda  }$ is symmetric in $\pm j$ and vanishes at $j=0$.  It is therefore also positive, $\psi_{  \lambda  }(j;z)\geq 0$. From the second line,
\begin{equation}\label{lo} L_{  \lambda  }(j;z) = -\frac{1}{2}\,j\cdot D^\dagger \id S_{  \lambda}(z) + \psi_{  \lambda  }(j;z) +\psi^\star_{  \lambda  }(D^\dagger \id S_{  \lambda}/2;z)
\end{equation}
which is \eqref{low}.

Next, we show that $\psi^\star_{  \lambda  }$ in \eqref{pst} is the same function as appears in \eqref{gger}. By replacing $\psi_{  \lambda  }$ in \eqref{lo} via \eqref{Lpsi} we find for \eqref{pst} the convex
\begin{equation}\label{Hpsistar1}
\psi^\star_{  \lambda  }(f;z) = H_{  \lambda  }(f-D^\dagger \id S_{  \lambda}/2;z)+L_{  \lambda  }(0;z)
\end{equation}
which is symmetric in $\pm f$, positive and vanishes only at $f=0$. Following the Hamilton equation in \eqref{zero cost flow}, the zero-cost flow must satisfy
\begin{equation}\label{pH2}
   j_{z, \lambda} =\partial_f H_{  \lambda  }(f;z)=\partial_f \psi^\star_{  \lambda  }(f + D^\dagger \id S_{  \lambda}/2;z) \qquad \text{ at } f = 0
\end{equation}
We conclude from \eqref{pH2} that the typical path $\dot{z} = D j_{z, \lambda}$ has indeed the generalized gradient flow structure \eqref{gger}. \\

  For illustration, the Hamiltonian and Lagrangian of Example \ref{cwex} appear in Appendix \ref{section curie weiss}, respectively equations \eqref{Hamiltonian two level system} and \eqref{lagrangian two level}. We refer to \cite{lagrangianmarkov, fren, largemarkov} for even more examples.

\section{Quasistatic gradient flow}\label{section quasistatic analysis}
We introduce an external protocol $ \lambda \to \lambda(\ve t)$ for $t \in [t_1 = \frac{\tau_1}{\ve}, t_2 = \frac{\tau_2}{\varepsilon}]$, which traces out a path in parameter space from an initial $\lambda_i = \lambda_{\tau_1}$ to the final $\lambda_f = \lambda_{\tau_2}$. The rate $\varepsilon > 0$ will go to zero in the quasistatic limit.  More specifically, the entropy $S_{  \lambda  } \rightarrow S_{\lambda(\ve t)}$ becomes time-dependent and, as a consequence,  the Lagrangian and the Hamiltonian become time-dependent as well, to govern the large deviation structure around the equation of motion,
\begin{align}
   \dot{z} & = D j_{z, \lambda(\varepsilon t)} = D \partial_f \psi_{  \lambda(\ve t)  }^\star\left(D^\dagger \id S_{\lambda(\varepsilon t)}/2;z \right) \label{quasistatic gradient flow eq} \\
   & \iff  L_{\lambda(\varepsilon t)}(j_{z,\lambda(\varepsilon t)}, z) = 0 \iff j_{z, \lambda(\varepsilon t)} = \partial_f H_{\lambda(\varepsilon t)}(f;z) \text{ and } f_{\lambda(\varepsilon t)} = 0 \nonumber
\end{align}
We denote the solution to this equation as $z_t^{(\varepsilon)}$ with fixed initial condition $z_{t = t_1}^{(\varepsilon)} = z^{*}_{\lambda(\tau_1)}$ corresponding to equilibrium at control value $\lambda_{\tau_1} =\lambda_i$ at time $t_1 = \tau_1/\ve$. We assume that everything is smooth in $\varepsilon$ to expand the solution $z_t^{(\varepsilon)}$ of \eqref{quasistatic gradient flow eq} around $\varepsilon = 0$,
\begin{align}\label{z epsilon expansion}
    z_t^{(\varepsilon)} = & z^{*}_{\lambda(\varepsilon t)} + \varepsilon \,\Delta z_{t}  + O(\varepsilon^2)
\end{align}
which will be needed to evaluate the action on shell.  The idea is to apply Noether's theorem for the path-space action for quasistatic gradient flow dynamics \eqref{quasistatic gradient flow eq} and show that the entropy is a Noether charge in both the Lagrangian and Hamiltonian formalism.\\

The leading-order term  $z^{*}_{\lambda(\varepsilon t)}$ corresponds to the reversible trajectory for control parameters $\lambda(\varepsilon t)$, {\it i.e.}, it is the instantaneous equilibrium condition:
\begin{equation}\label{fixed point z}
    0 = \dot{z}^*_\lambda = D j_{z^*_\lambda} = D \partial_f \psi^\star_{  \lambda  }(D^\dagger \id S^*_{\lambda}/2;z^*_\lambda) \iff L_{  \lambda  }(j_{z^*_\lambda}, z^*_\lambda) = 0 
\end{equation}
Hence, the current $j_{z^*_\lambda} = $ constant.  Detailed balance \eqref{dbe} further implies
\begin{equation*}
     L_{  \lambda  }(- j_{z^*_\lambda}, z^*_\lambda) =  L_{  \lambda  }(j_{z^*_\lambda}, z^*_\lambda) + \dot{z}_\lambda^* \id S^*_{\lambda} =  L_{  \lambda  }(j_{z^*_\lambda}, z^*_\lambda) = 0
     \end{equation*}
     Remembering that $L_{  \lambda  }(j;z)$ has a unique minimum at $j = j_{z, \lambda}$ it follows that $- j_{z^*_\lambda} = j_{z^*_\lambda}$ or $j_{z^*_\lambda} = 0$. Moreover,
\begin{equation}\label{dS on shell}
  0 =  j_{z^*_\lambda} = \partial_f \psi^\star_{  \lambda  }(D^\dagger \id S^*_{\lambda}/2;z^*_\lambda) \Longrightarrow D^{\dagger} \id S^*_\lambda = 0
\end{equation}
where the last line follows from \eqref{cond:psi*-min-at-zero}.\\ 
 
The same result can also be obtained in the Hamiltonian formalism since detailed balance \eqref{dbham} with $\frac{\id S_\lambda}{\id z}(z_\lambda^*) = 0$ yields $ H_{  \lambda  }(f;z^*_\lambda) = H_{  \lambda  }(-f;z^*_\lambda)$ and thus
\begin{equation*}
\dot{z}_\lambda^* = \partial_f H_{\lambda}(0;z_{\lambda}^*) = - \partial_f H_{\lambda}(0;z_{\lambda}^*) = - \dot{z}^*_\lambda
\end{equation*}
or $\dot{z}_\lambda^* = 0$, as in the Lagrangian case. Consequently, the detailed balance conditions \eqref{dbe}-\eqref{dbham} indeed imply the time-reversal invariance of the equilibrium state. \\

The system lags behind this instantaneous equilibrium state, as described by the higher-order terms in \eqref{z epsilon expansion} with $\Delta z_{t}$ being its first-order correction. We only require that $\Delta z_{t}$ (the coefficient for the linear term in $\ve$ as defined in \eqref{z epsilon expansion}) remains bounded uniformly as $\ve \downarrow 0$.  We refer to Appendix \ref{section curie weiss} and the explicit solutions \eqref{delta m hamiltonian}--\eqref{delta p hamiltoniaan} in the case of a quasistatic Curie-Weiss dynamics. \\
\\
 
Note that along a general trajectory $(z_t, j(t))$
\begin{equation}\label{zdot ds}
\frac{\id}{\id t} S_{\lambda(\ve t)}(z_t) = \dot{z} \cdot \id S_{\lambda(\ve t)} + \ve  \dot{\lambda}(\ve t) \cdot \partial_\lambda S_{\lambda(\ve t)}(z_t)
\end{equation}
which, for the trajectories \eqref{z epsilon expansion}, reduces to
\begin{equation}\label{ine}
\frac{\id}{\id t} S_{\lambda(\ve t)} \left(z_t^{(\ve)} \right) =\ve \partial_\lambda{z^*_{\lambda(\ve t)}} \dot{\lambda}(\ve t) \cdot \id S_{\lambda(\ve t)}(z^*_{\lambda(\ve t)}) + \ve  \dot{\lambda}(\ve t) \cdot \partial_\lambda S_{\lambda(\ve t)}(z_{\lambda(\ve t)}^*) + O(\ve^2) = O(\ve^2)
\end{equation}
where, in the last line, we used the equilibrium conditions \eqref{eq condition}. In other words, as correct to quadratic order in $\ve$ there is a constant entropy for the trajectories \eqref{z epsilon expansion}.  The objective of the present paper is to relate that to Noether's theorem.

\section{Entropy as a Noether charge}\label{section symmetry}

\subsection{Hamiltonian formalism}
We start with the expression of the path-space action in terms of the Hamiltonian,
\begin{equation}
    \mathcal{A} = - S_{ \lambda_1  }(z_{t_1}) + \int_{t_1}^{t_2} \id t \ \big[f_t \cdot j(f_t;z_t) - H_{\lambda(\varepsilon t)}(f_t;z_t) \big]
\end{equation}
where the thermodynamic force $f$ and the macroscopic condition $z$ are independent variables.  We have inserted the quasistatic protocol through $\lambda = \lambda(\varepsilon t)$ for  $\ve \downarrow 0$.\\ 
We consider the \textit{continuous} symmetry, exactly like in \eqref{onslow}--\eqref{f shift onsager}, 
\begin{equation}\label{shift df}
 t \to t' = t, \qquad   z_t \to z'_t = z_t, \qquad f_t \to f'_t = f_t + \eta \ D^{\dagger} \id S_{\lambda(\varepsilon t)}(z_t)
\end{equation}
with $S_{\lambda(\varepsilon t)}$ from \eqref{gger}. 
Under the shift \eqref{shift df}, the action changes as
 
\begin{align}
    \delta \mathcal{A}  
    &= \eta \int_{t_1}^{t_2} \id t  \Big[ D j(t) \cdot \id S_{\lambda(\varepsilon t)}(z_t) - \partial_f H_{\lambda(\varepsilon t)}(f_t;z_t) \cdot D^{\dagger} \id S_{\lambda(\varepsilon t)}(z_t) \Big] \nonumber \\
    &= \eta \int_{t_1}^{t_2} \id t  \Bigg[\frac{\id S_{\lambda(\varepsilon t)}}{\id t}(z_t) - \ve \ \dot{\lambda}(\ve t) \cdot \partial_\lambda S_{\lambda(\ve t)}(z_t) - \partial_f H_{\lambda(\varepsilon t)}(f_t;z_t) \cdot D^{\dagger} \id S_{\lambda(\varepsilon t)}(z_t) \Bigg] \nonumber 
\end{align}
where we have used \eqref{zdot ds}.  One should keep in mind that $t_1 = \tau_1/\ve, t_2 = \tau_2/\ve$.  Note that $\delta \mathcal{A}$ does not form a total derivative $\id \zeta/\id t$ for general $(z_t, f_t)$. Focussing instead on quasistatic trajectories \eqref{z epsilon expansion} only, it follows that
\begin{align*}
\ve \ \dot{\lambda}(\ve t) \cdot \partial_\lambda S_{\lambda(\varepsilon t)} \left(z_t^{(\ve)} \right) & = \ve \ \dot{\lambda}(\ve t) \cdot \partial_\lambda S_{\lambda(\varepsilon t)}(z^*_{\lambda(\ve t)}) + \ve^2 \dot{\lambda}(\ve t) \cdot \partial^2_\lambda S_{\lambda(\ve t)}(z^*_{\lambda(\ve t)}) \Delta z_{\ve t}  + O(\ve^3) = O(\ve^2) \nonumber \\
   & \nonumber \\
   \partial_f H_{\lambda(\varepsilon t)}\left(f_t^{(\varepsilon)};z_t^{(\varepsilon)} \right) 
  & = j_{z^*_{\lambda(\varepsilon t)}} +  \varepsilon \left(\partial_z \partial_f H_{\lambda(\varepsilon t)}(0;z^*_{\lambda(\varepsilon t)}) \Delta z_{\varepsilon t} \right) +  O(\varepsilon^2) = O(\ve) \\
     & \nonumber \\
 D^{\dagger} \id S_{\lambda(\varepsilon t)}\left(z_t^{(\varepsilon)} \right) & = D^{\dagger} \id S_{\lambda(\varepsilon t)}(z^*_{\lambda(\varepsilon t)} ) + \varepsilon  D^{\dagger} \left( \id^2 S_{\lambda(\varepsilon t)}(z^*_{\lambda(\varepsilon t)}) \Delta z_{\varepsilon t} \right) + O(\varepsilon^2) = O(\ve)
\end{align*}
where we have used the equilibrium conditions \eqref{eq condition},  $\id S_{\lambda}^* = 0, \partial_\lambda S_\lambda^* = 0$ and $j_{z_\lambda^*}  = 0$. The change in the action then becomes
\begin{equation}
     \delta \mathcal{A} = \eta \int_{t_1}^{t_2} \id t  \left[ \frac{\id S_{\lambda(\varepsilon t)}}{\id t}(z_t^{(\ve)}) + O(\varepsilon^2) \right] = \eta \left( S_{\lambda(\ve t_2)}(z_{t_2}^{(\ve)}) - S_{\lambda(\ve t_1)}(z_{t_1}^{(\ve)}) \right) + \eta \ O(\ve)
\end{equation}
where the $O(\ve^2)$ terms inside the integral are assumed to remain bounded over the entire trajectory. Expanding the entropy terms further in $\ve$ yields
\begin{align*}
     \delta \mathcal{A} & = \eta \left(S_{\lambda(\ve t_2)}(z^*_{\lambda(\ve t_2)}) - S_{\lambda(\ve t_1)}(z^*_{\lambda(\ve t_1)}) \right) + \eta \ve \left( \id S_{\lambda(\ve t_2)}(z^*_{\lambda(\ve t_2)}) \Delta z_{\ve t_2} - \id S_{\lambda(\ve t_1)}(z^*_{\lambda(\ve t_1)}) \Delta z_{\ve t_1} \right) + \eta \ O(\ve) \\
     & = \eta \left(S_{\lambda(\ve t_2)}(z^*_{\lambda(\ve t_2)}) - S_{\lambda(\ve t_1)}(z^*_{\lambda(\ve t_1)}) \right) + \eta \ O(\ve) \\
     & = \eta \int_{\tau_1}^{\tau_2} \id \tau  \frac{\id S_{\lambda(\tau)}}{\id \tau}(z^*_{\lambda(\tau)}) + \eta \ O(\ve)
\end{align*}
where we have used $\id S_{\lambda(\ve t)}(z^*_{\lambda(\ve t)}) = 0$ in the middle. Taking the quasistatic limit $\ve \downarrow 0$, this reduces to 
\begin{align}\label{delta A ds/dt}
    \lim_{\ve \downarrow 0} \delta \mathcal{A} =  \eta \int_{\tau_1}^{\tau_2} \id \tau  \frac{\id S_{\lambda(\tau)}}{\id \tau}(z^*_{\lambda(\tau)}) = \eta \left(S_{\lambda(\tau_2)}(z^*_{\lambda(\tau_2)}) - S_{\lambda(\tau_1)}(z^*_{\lambda(\tau_1)}) \right) 
\end{align}
which is a geometric expression, {\it i.e.}, independent on time reparametrizations.
 
That has the form of a quasisymmetry, \cite{quasisymmetry}:
\begin{equation}\label{quasisymmetry}
    \delta \mathcal{A} = \eta   \int_{\tau_1}^{\tau_2} \id \tau \  \frac{\id \zeta}{\id \tau},   
    \qquad \zeta = S_{\lambda}
\end{equation}
On the other hand, under the coordinate transform $  f_t \to f_t + \delta f_t, z_t \to z_t  $, the action changes as
\begin{equation}\label{general change hamiltonian df shift}
    \delta \mathcal{A} = \eta \int_{t_1}^{t_2} \id t \  \delta f_t \left(  j(t)   - \partial_f H_{  \lambda(\ve t)  }(f_t; z_t)\right) 
\end{equation}
  As explained in Section \ref{subsection lagrangian and hamiltonian}, the gradient flow also satisfies $j_z = \partial_f H_\lambda$ \textit{on shell}, leading to the stationarity of the action $\delta \mathcal{A} |_{\text{on shell}} = 0$. Consequently, from \eqref{delta A ds/dt}, the entropy is   conserved in the quasistatic limit $\ve \downarrow 0$:
\begin{equation*}
    S_{\lambda_f} = S_{\lambda_i}
\end{equation*}

In summary, we have shown that the entropy function $S_{  \lambda  }$ is a constant of motion for the shift \eqref{shift df} in the quasistatic limit. Recall however that we have used the expression `entropy' for more general thermodynamic potentials (free energies).  Therefore, its invariance means different things depending on the actual physical situation and what variables are kept constant or not during the quasistatic evolution.  For example, the state variable $z$ can be the volume of a box with thermally isolated walls and containing a dilute gas in contact with a pressure bath.  The protocol would then make a quasistatic change in the (external) pressure $p$ and we would have $S_p(V)$ to represent the thermodynamic potential as a function of the enthalpy $H(V,p)$ and the pressure $p$.  The result is then that $S_p(V)$ is a Noether charge.  That was the example treated in \cite{langevin_noether}. 

\subsection{Lagrangian formalism}
We can also use the Lagrangian setup as the starting point for a continuous symmetry.  We write the action
\begin{equation}
    \mathcal{A} = - S_{ \lambda_1  }(z^*_{\lambda_1}) +  \int_{t_1}^{t_2} \id t \ L_{\lambda(\varepsilon t)}(j(t); z_t)
\end{equation}
which is a function of the trajectory starting in equilibrium state $z^*_{\lambda_1}$.
We introduce a quasistatic dynamics through $\lambda = \lambda(\varepsilon t)$ for some $\ve \downarrow 0$. We consider the \textit{continuous} symmetry, exactly like in \eqref{j shift onsager},  
\begin{equation}\label{shift dj}
   t \to t' = t, \qquad   z_t \to z_t' = z_t, \qquad j(t) \to j'(t) = j(t) - 2 \eta j(t)  
\end{equation}
with $S_{\lambda(\varepsilon t)}$ from \eqref{gger} and small $\eta$. 
Under the shift \eqref{shift dj}, the action changes as
 
\begin{align*}
    \delta \mathcal{A} &= -2 \eta\int_{t_1}^{t_2} \id t \ j(t) \cdot \frac{\partial L_{\lambda(\ve t)}}{\partial j}(j(t); z_t) \\
    & = \eta \int_{t_1}^{t_2} \id t \Big[ j(t) \cdot D^{\dagger} \id S_{\lambda(\ve t)}(z_t) -2 j(t) \cdot \partial_j \psi_{\lambda(\ve t)  }(j(t),z_t) \Big] \\
    &= \eta \int_{t_1}^{t_2} \id t \Bigg[ \frac{\id S_{\lambda(\ve t)}}{\id t}(z_t) - \ve \ \dot{\lambda}(\ve t) \cdot \partial_\lambda S_{\lambda(\ve t)}(z_t)  -2 j(t) \cdot \partial_j \psi_{\lambda(\ve t)  }(j(t),z_t)\Bigg ]
\end{align*}
where we used \eqref{lo} in the middle line and \eqref{zdot ds} in the last. This $\delta \mathcal{A}$  does not form a total derivative $\id \zeta/\id t$ for general $(z_t, j(t))$. Focussing instead on quasistatic trajectories \eqref{z epsilon expansion} only, $z_t^{(\varepsilon)} = z^*_{\lambda(\varepsilon t)} + \varepsilon \Delta z_{\varepsilon t} + O(\varepsilon^2)$ and $j^{(\ve)}(t) = j_{z_{\lambda(\ve t)}^*} + \ve \Delta j(t) + O(\ve^2)$, it follows that
\begin{align*}
\ve \ \dot{\lambda}(\ve t) \cdot \partial_\lambda S_{\lambda(\varepsilon t)} \left(z_t^{(\ve)} \right) & = \ve \ \dot{\lambda}(\ve t) \cdot \partial_\lambda S_{\lambda(\varepsilon t)}(z^*_{\lambda(\ve t)}) + \ve^2 \dot{\lambda}(\ve t) \cdot \partial^2_\lambda S_{\lambda(\ve t)}(z^*_{\lambda(\ve t)}) \Delta z_{\ve t}  + O(\ve^3) = O(\ve^2) \nonumber \\
   & \nonumber \\
    - 2 j^{(\ve)}(t) \cdot \partial_j \psi_{\lambda(\ve t)  } \left(j^{(\ve)}(t),z_t^{(\ve)} \right) &= \ve \Delta j_{\ve t} \cdot  \partial_j \psi_{ \lambda(\ve t)  }(0, z_{\lambda(\ve t)}^*) \\
    & + \ve^2 \Delta j_{\ve t} \cdot \left(\partial_j^2 \psi_{ \lambda(\ve t)  }(0, z^*_{\lambda(\ve t)}) \Delta j_{\ve t} + \partial_j \partial_z \psi_{ \lambda(\ve t)  }(0, z^*_{\lambda(\ve t)}) \Delta z_{\ve t} \right) + O(\ve^3) \\
    & = O(\ve^2)
\end{align*}
where we used the equilibrium conditions \eqref{eq condition} and $\partial_j \psi_{ \lambda(\ve t)  }(0,z^*_{\lambda(\ve t)}) = 0$ since $\psi_\lambda$ reaches its minimum at $j = 0$ and is strictly convex. Moreover, the $O(\ve^2)$ terms are assumed to remain bounded over the entire trajectory.  The change in the action then becomes
\begin{equation}
     \delta \mathcal{A} = \eta \int_{t_1}^{t_2} \id t  \left[ \frac{\id S_{\lambda(\varepsilon t)}}{\id t}(z_t^{(\ve)}) + O(\varepsilon^2) \right] = \eta \left( S_{\lambda(\ve t_2)}(z_{t_2}^{(\ve)}) - S_{\lambda(\ve t_1)}(z_{t_1}^{(\ve)}) \right) + \eta \ O(\ve)
\end{equation}
since $t_1 = \tau_1/\ve, t_2 = \tau_2/\ve$. Expanding the entropy terms further in $\ve$ yields
\begin{align*}
     \delta \mathcal{A} & = \eta \left(S_{\lambda(\ve t_2)}(z^*_{\lambda(\ve t_2)}) - S_{\lambda(\ve t_1)}(z^*_{\lambda(\ve t_1)}) \right) + \eta \ve \left( \id S_{\lambda(\ve t_2)}(z^*_{\lambda(\ve t_2)}) \Delta z_{\ve t_2} - \id S_{\lambda(\ve t_1)}(z^*_{\lambda(\ve t_1)}) \Delta z_{\ve t_1} \right) + \eta \ O(\ve) \\
     & = \eta \left(S_{\lambda(\ve t_2)}(z^*_{\lambda(\ve t_2)}) - S_{\lambda(\ve t_1)}(z^*_{\lambda(\ve t_1)}) \right) + \eta \ O(\ve) \\
     & = \eta \int_{\tau_1}^{\tau_2} \id \tau  \frac{\id S_{\lambda(\tau)}}{\id \tau}(z^*_{\lambda(\tau)}) + \eta \ O(\ve)
\end{align*}
where we have used $\id S_{\lambda(\ve t)}(z^*_{\lambda(\ve t)}) = 0$ in the middle. Taking the quasistatic limit $\ve \downarrow 0$, this reduces to 
\begin{align}\label{delta a dsdt 2}
    \lim_{\ve \downarrow 0} \delta \mathcal{A} =  \eta \int_{\tau_1}^{\tau_2} \id \tau  \frac{\id S_{\lambda(\tau)}}{\id \tau}(z^*_{\lambda(\tau)}) = \eta \left(S_{\lambda(\tau_2)}(z^*_{\lambda(\tau_2)}) - S_{\lambda(\tau_1)}(z^*_{\lambda(\tau_1)}) \right)
\end{align}
which is a geometric expression, {\it i.e.}, independent on time reparametrisations, and has the form of a quasisymmetry \eqref{quasisymmetry}.  
On the other hand, under the transform $  j(t) \to j(t) + \delta j(t), z_t \to z_t  $, the action changes as
\begin{equation}\label{general change lagrangian dj shift}
    \delta \mathcal{A} = \eta \int_{t_1}^{t_2} \id t \   \delta j(t) \cdot \frac{\partial L_{  \lambda(\ve t)  }}{\partial j}  (j(t); z_t)  
\end{equation}
which is zero \textit{on shell} because $\frac{\partial L_{  \lambda(\ve t)  }}{\partial j}(j_{z_t^{(\ve)}}; z_t^{(\ve)}) =0$, as explained in Section \eqref{subsection lagrangian and hamiltonian},   leading to the stationarity of the action $\delta \mathcal{A}|_{\text{on shell}} = 0$.    Therefore, from \eqref{delta a dsdt 2}, the Noether charge for the transformation \eqref{shift dj} is just the entropy $Q = S_\lambda$ and it is conserved in the quasistatic limit $\ve \downarrow 0$:
\begin{equation*}
    S_{\lambda_f} = S_{\lambda_i}
\end{equation*}

\section{Conclusion}
We have extended the relation between the entropy change $\Delta S_{  \lambda  }$ and the (discrete) time-reversal transformation to a continuous symmetry for quasistatic gradient dynamics. That generalizes ideas and results first expressed and obtained by Sasa \textit{et al.}, connecting the first part of the Clausius heat theorem with the Noether theorem.\\  The novelty of our approach is in considering the general large deviation structure for gradient flows, including the case of jump processes (leading to generalized gradient flow). That allows us to extend the results to pure relaxational processes including spin-flip or (chemical) reaction dynamics. We have illustrated that with Curie-Weiss and urn models.\\
The large deviation approach also sheds new light on the continuous symmetry: in the Hamiltonian version of the Noether argument, it is the shift in thermodynamic force that realizes the symmetry.  That suggests the picture where the probability of macroscopic trajectories is parametrized by the time-symmetric and the time-antisymmetric part in the action. That antisymmetric part is governed by a dissipation (or Onsager) functional and basically multiplies currents (fluxes) with (thermodynamic) forces. Time reversal is implemented by a change in the force while keeping the time-symmetric part unchanged.  We suggest seeing it as the analogue of a rotation in the two-dimensional plane to implement reflection (on the one-dimensional axis).\\
We believe that the results may prove interesting for {\it unifying} classical and quantum mechanical, stochastic, and general relativistic arguments for the existence of entropy as exact differential in reversible transformation, invariant under adiabatic ({\it i.e.}, no heat) conditions.  That has no relation however with the Second Law (non-decreasing entropy) which here is assumed from the outset in the very framework of gradient flow. \\
\\
\textbf{Acknowledgment:} The authors thank Shin-ichi Sasa for private correspondence.   Aaron Beyen is supported by the Research Foundation - Flanders (FWO) doctoral fellowship 1152725N.  

\appendix

\section{Curie-Weiss dynamics II}\label{section curie weiss}
As a final illustration, we continue with the kinetic Ising model on the complete graph, \cite{Kochma_ski_2013, friedli_velenik_2017},   with energy function  
\begin{equation}\label{energy and psi}
    E_{  \lambda  }(\sigma) := -  N \phi_{  \lambda  }(m^N(\sigma))
\end{equation}
only depending on the magnetization (function), $m^{N}(\sigma) :=\frac{1}{N} \sum_{i=1}^{N} \sigma_{i}$, with $\phi_{  \lambda  }(m)$ a smooth function whose exact form is irrelevant in what follows. We take the (finite) Curie-Weiss dynamics to be the Markov process on $K_N:= \{-1,+1\}^N$ with rates $c_{  \lambda  }(\sigma,k)$ for flipping $\sigma = (\sigma_{1}, \sigma_{2},..., \sigma_{k},...,\sigma_{N}) \to \sigma^k =  (\sigma_{1}, \sigma_{2},..., -\sigma_{k},...,\sigma_{N})$, given by
\begin{eqnarray}\label{sumr}
 c_{  \lambda  }(\sigma,k) &=&  \nu(\beta)\, e^{\frac{\beta}{2}(E_{  \lambda  }(\sigma)-E_{  \lambda  }(\sigma^k))} \label{alra} 
\end{eqnarray}
where $\beta$ represents the inverse temperature of the  thermal bath \footnote{  The temperature can also be considered as a control parameter $\lambda$.  } and $\nu = \nu(\beta)$ is a characteristic flip frequency possibly depending on $\beta$.   These   spin-flip rates satisfy detailed balance with respect to the potential \eqref{energy and psi}.
\\The backward generator $\cal {L}_{  \lambda  }^{N}$ of that Markov process  is
\begin{align*}
   \cal  {L}^N_{  \lambda  } f(\sigma) &= \sum_{k = 1}^N \left[a_{  \lambda  }(m^N(\sigma))+ b_{  \lambda  }(m^N(\sigma))\sigma_k) \right] \left( f(\sigma^k) - f(\sigma)\right) \\
    a_{  \lambda  }(m^N(\sigma)) &= \nu  \cosh\left(\beta \phi_{  \lambda  }'(m^N(\sigma)) \right) + O \left( \frac{1}{N}\right) \\
    b_{  \lambda  }(m^N(\sigma)) &= -\nu \sinh \left(\beta \phi_{  \lambda  }'(m^N(\sigma)) \right) + O \left( \frac{1}{N}\right)
\end{align*}
To simplify the algebra, we use the $(a_{  \lambda  },b_{  \lambda  })$ parameterization as long as possible, but we remain in the Curie-Weiss model at all times. Note in particular that $a_{  \lambda  }(m) > b_{  \lambda  }(m)$ for all $m$.\\
The Hamiltonian equals
\begin{align}\label{Hamiltonian two level system}    H_{  \lambda  }(f;m) & = \sinh(2f) \Big[(a_{  \lambda  }(m) +  m \ b_{  \lambda  }(m)) \tanh(f) - (b_{  \lambda  }(m) + m \ a_{  \lambda  }(m)) \Big]
\end{align}
See {\it e.g.}, p109 in \cite{largemarkov}.
For the Lagrangian $L(\dot{m}; m)$, we put $\dot m = j$, and we maximize $f \dot{m} - H_{  \lambda  }(f;m)$   with respect to $f$,  
\begin{equation}\label{p}
  0 = \dot{m} - \frac{\partial H_{ \lambda  }}{\partial f}(f; m)  \Longrightarrow f_{  \lambda  }(\dot{m}; m) = \frac{1}{2} \log \left(\frac{\pm \sqrt{\dot{m}^2+4 \left(1-m^2\right) (a_{  \lambda  }^2-b_{  \lambda  }^2)} + \dot{m}}{2 (1-m) (a_{  \lambda  }-b_{  \lambda  })}\right) 
\end{equation}
where we need the $+$ sign to get a maximum since then
\[
    \frac{\partial^2 H_{  \lambda  }}{\partial f^2}(f_{  \lambda  }(\dot{m}; m), m) = 2 \sqrt{\dot{m}^2+ 4 \left(1-m^2\right) (a_{  \lambda  }^2-b_{  \lambda  }^2)} > 0
\]
We have used that $-1 \leq m \leq 1$ and $a_{  \lambda  }(m) > b_{  \lambda  }(m)$; in fact $a_{  \lambda  }^2-b_{  \lambda  }^2 = \nu^2$. The corresponding Lagrangian becomes
\begin{align}\label{lagrangian two level}
  \hspace{-1 cm} L_{  \lambda  }(m, \dot{m}) &= \frac{1}{2} \Bigg[-\sqrt{\dot{m}^2+4 \left(1-m^2\right)\nu^2}+\dot{m} \log \left(\frac{\sqrt{\dot{m}^2+4 \left(1-m^2\right) \nu^2 }+\dot{m}}{2 (1-m) (a_{  \lambda  }-b_{  \lambda  })}\right) +2( a_{  \lambda  }+  b_{  \lambda  } m)\Bigg]
\end{align}
  Moreover, one readily checks that the functions $\psi_{  \lambda  }, \psi^\star_{  \lambda  }$ from Section \ref{canonical structure} are given by
\begin{align*}
  \psi_{  \lambda  }(\dot{m};m) &= \frac{1}{2} \left[2 \sqrt{1-m^2} \nu  - \sqrt{\dot{m}^2 + 4 (1-m^2) \nu^2} + \dot{m} \log \left(\frac{\dot{m} + \sqrt{\dot{m}^2 + 4 (1-m^2) \nu^2}}{2 \nu \sqrt{1-m^2}} \right)  \right] \\
   \psi^\star_{  \lambda  }(f,m) &= \nu \sqrt{1-m^2} \left(\cosh(2 f)-1 \right) 
\end{align*}
 
The macroscopic equation of motion is obtained from $L_{  \lambda  }(m, \dot{m}) = 0$, demanding that the logarithmic term and the other terms cancel separately:
\begin{equation}\label{mdot = g curie weiss}
    \dot{m} = G_{  \lambda  }(m), \qquad \text{ for }\; G_{  \lambda  }(m) = - 2 (b_{  \lambda  }(m) + a_{  \lambda  }(m) m)
\end{equation}
which agrees with \cite{fiori2024specificheatdrivencurieweiss, beyen2024phase}.
The equilibrium solutions follow from $\dot{m}^*_{  \lambda  } = 0$
\begin{equation}\label{equilibrium cw}
    G_{  \lambda  }(m^*_{  \lambda  }) = 0 \iff m^*_{  \lambda  } = -\frac{b_{  \lambda  }(m^*_{  \lambda  })}{a_{  \lambda  }(m^*_{  \lambda  })} = \tanh( \beta \phi_{  \lambda  }'(m^*_{  \lambda  }))
\end{equation}
As in \eqref{dbe}, for the free energy in \eqref{cw m dynamics}, we have detailed balance
\begin{equation}\label{antisymmetric part lagrangian}
    L_{  \lambda  }(m,\dot{m}) - L_{  \lambda  }(m,-\dot{m}) = \dot{m} \left(\arctanh(m) + \frac{1}{2} \log \left(\frac{a_{  \lambda  }+b_{  \lambda  }}{a_{  \lambda  }-b_{  \lambda  }} \right) \right) 
    = \dot{m} \partial_m \mathcal{F}_{  \lambda  } = \frac{\id \mathcal{F}_{  \lambda  }}{\id t} 
\end{equation}
and $\mathcal{F}_{  \lambda  }$ solves the stationary Hamilton-Jacobi equation,
\begin{equation}\label{h partial s = 0}    H_{  \lambda  }(m, \partial_m \mathcal{F}_{  \lambda  }) =0
\end{equation}

As in Section \ref{section quasistatic analysis}, we introduce a protocol $ \lambda \to  \lambda(\varepsilon t)$ for $t \in [t_1,t_2] = [\frac{\tau_1}{\ve}, \frac{\tau_2}{\varepsilon}], \ve>0$, with $\ve \downarrow 0$ making the process quasistatic. To be specific, we take a slowly-varying energy function $\phi_\lambda \to \phi_{  \lambda(\varepsilon t)  }$ or equivalently  $a_\lambda \to a_{  \lambda(\varepsilon t)  }$, and $b_\lambda \to b_{  \lambda(\varepsilon t)  }$, {\it e.g.} due to a time-dependent magnetic field $h \to h(\epsilon t)$,
giving rise to the slowly varying Hamiltonian
\begin{align}\label{curie weiss h(m,p,t)}
    H_{  \lambda(\ve t)}(f;m) & = \sinh(2f) \left[ (a_{  \lambda(\varepsilon t)  } + b_{  \lambda(\varepsilon t)  } m) \tanh(f) -(b_{  \lambda(\varepsilon t)  } + a_{  \lambda(\varepsilon t)  } m) \right]
\end{align}
with Hamilton equations of motion
\begin{align*}
    \dot{m} & = \frac{\partial H_{  \lambda(\ve t)}}{\partial f} =  2 \sinh{(2f)} \big(a_{  \lambda(\varepsilon t)  } + m b_{  \lambda(\varepsilon t)  } \big) - 2 \cosh{(2f)} \big(b_{  \lambda(\varepsilon t)  } + m a_{  \lambda(\varepsilon t)  } \big)\\
    \dot{f} & = - \frac{\partial H_{  \lambda(\ve t)}}{\partial m} = \sinh{(2f)} \Big[a_{  \lambda(\varepsilon t)  } + m a'_{  \lambda(\varepsilon t)  } + b'_{\lambda(\varepsilon t)} - \tanh(f) \big(b_{  \lambda(\varepsilon t)  } + a'_{  \lambda(\varepsilon t)  } + m b'_{  \lambda(\varepsilon t)  } \big) \Big] 
\end{align*}
We denote the solutions to these equations as $m_t^{(\varepsilon)}, f_t^{(\varepsilon)}$ with initial conditions $m_{t = t_1}^{(\varepsilon)} = m^{*}_{\lambda_1}$ and $f_{t = t_1}^{(\varepsilon)} = f^{*}_{\lambda_1} = 0$. Perturbatively in $\varepsilon$,
\begin{align*}
    m_t^{(\varepsilon)} & = m^{*}_{\lambda(\varepsilon t)} + \varepsilon \Delta m_{t} + O(\varepsilon^2) \\
   f_t^{(\varepsilon)} & = f^{*}_{\lambda(\varepsilon t)} + \varepsilon \Delta f_{t} + O(\varepsilon^2)
\end{align*}
Here, the leading order term $O(\varepsilon^0)$ corresponds to the equilibrium state $m^*_\lambda$ but with $\lambda$ replaced by $\lambda(\varepsilon t)$, {\it i.e.}, an instantaneous equilibrium state.  Up to order $O(\varepsilon^2)$ we find
\begin{align}\label{delta m hamiltonian}
   \Delta m_t = & \frac{ \dot{\lambda}(\tau) \partial_\lambda m^*_{\lambda(\tau)}}{\partial_m G_{  \lambda(\tau)  }(m_{\lambda(\tau)}^*)},\quad \text{  with  }\;     \partial_\lambda m^* = - \frac{\partial_\lambda(b_\lambda/a_\lambda)}{1 + \partial_m(b_\lambda/a_\lambda)}  \\
    \Delta f_t = &\  0
    \label{delta p hamiltoniaan}
\end{align}
with $G_\lambda$ from \eqref{mdot = g curie weiss} and we have used that $f^*_\lambda = 0$ and thus also $\partial_\lambda f^*_\lambda = 0$. Note that $\Delta m_t$ remains bounded only when 
$\partial_m G_\lambda(m_{\lambda(\tau)}^*)$ does not diverge, which means we are not allowed to pass the critical point.

\bibliographystyle{unsrt}
\bibliography{bib.bib}

\end{document}